# Demonstration of room-temperature, magnetic-field-free, and ultralow-power spin-orbit torque logic device based on composition-uniform van der Waals magnet $Fe_3GaTe_2$

Zhengxiao Li[1,2], Qianbiao Liu[1], Wenliang Zhu,[3] and Lijun Zhu[1,2*]

1. *State Key Laboratory of Semiconductor Physics and Chip Technologies, Institute of Semiconductors, Chinese Academy of Sciences, Beijing 100083, China*
2. *Center of Materials Science and Optoelectronics Engineering, University of Chinese Academy of Sciences, Beijing 100049, China*
3. *College of Physics and Information Technology, Shaanxi Normal University, Xi'an 710062, China*

*ljzhu@semi.ac.cn

**Van der Waals magnets have received blooming interest in material science and spintronics for the advantages of low magnetization, flexible stacking, strong tunability, and dangling-bond-free interfaces. However, there has been no report of room-temperature magnetic-field-free spin logic device based on a van der Waals magnet. Here, we demonstrate a room-temperature, field-free, low-power, scalable spin-orbit torque in-memory computing logic device utilizing a van der Waals magnet $Fe_3GaTe_2$ bit with strong perpendicular magnetic anisotropy and a Pt channel with strong spin Hall effect, high electrical conductivity, and electric asymmetries. Current pulse width-dependent switching measurement reveals a low intrinsic critical magnetic-field-free switching current of $1.8\times10^7$ A/cm², a high thermal stability factor of 50, and the potential of an ultralow write power of 1.5 fJ/bit at 1 ns pulse width for a typical industrial-level device dimensions (40 times lower than the industrially optimized W/FeCoB device). These results suggest a great potential of van der Waals magnets in dense, ultralow-power, scalable in-memory computing technologies.**

**Introduction.** Artificial intelligence technologies create rapidly increasing computing demand that can hardly be fulfilled by the traditional charge-based von Neumann computing[1]. In-memory computing spin logic is of great interest for the development of high-performance, non-Von Neumann chips. Despite the efforts over two decades, the study of spin logic has been mainly focused on conventional metallic magnets[2-8]. Recently, van der Waals (vdw) magnets have received blooming interest in material science and spintronics for the advantages of low magnetization, flexible stacking, strong tunability, and dangling-bond-free interfaces[9-15]. Of particular interest is the vdw ferromagnet $Fe_3GaTe_2$ with above-room-temperature Curie temperature,[16] strong perpendicular magnetic anisotropy (PMA), giant anomalous Hall effect[12], and large tunnel magnetoresistance[17-19].

So far, magnetic-field-free switching of PMA $Fe_3GaTe_2$ has been reported based on side-wall Pt capping[13], interfacial coupling of antiferromagnet (MnPt)[14], and resistive low-symmetry crystals ($TaIrTe_4$ or $WTe_2$).[20-25] Switching of $Fe_3GaTe_2$ has also been reported by the combination of Fe-concentration gradient and an in-plane magnetic anisotropy preset by a magnetic field of 20 mT.[26] However, there has been no report of field-free switching of a uniform, regularly patterned $Fe_3GaTe_2$ device based on high-conductivity spin Hall metals like Pt. Note that large-scale integration of spintronic memory and logic devices requires uniform, regular patterning of the magnetic bits to assure the wafer-scale identicalness, while a high-conductivity spin Hall metal is highly beneficial for the enhancement of the power efficiency and the endurance of the SOT devices.[27] Moreover, there has been no report on the thermal stability factor ($\Delta$) and the critical switching current ($j_{c0}$) of uniform $Fe_3GaTe_2$ from pulse-width dependent measurement despite the switching current density ($j_c$) and data retention decreases strongly at low $\Delta$ values and cannot reflect the switching performance of fast operations in logic and memory applications.[28]

There has also been no report of room-temperature magnetic-field-free spin logic device based on a uniform vdw magnet. Previous $Fe_3GeTe_2$ device[29] had a working temperature of < 190 K, while the $Fe_4GeTe_2$ devices[30] require a Fe-concentration gradient and a high magnetic field of up to 1 T to induce in-plane magnetic anisotropy.

In this work, we report the first demonstration of a low-power, scalable, room-temperature, magnetic-field-free SOT Boolean logic based on composition-uniform etching-defined $Fe_3GaTe_2$/Pt devices with strong PMA but no composition gradient (Fig. 1a). The field-free switching is achieved by electric asymmetries within the high-conductivity spin Hall metal Pt.

**Room temperature field-free switching.** To obtain a strong PMA and a high $\Delta$, a 37 nm-thick $Fe_3GaTe_2$ layer is first mechanically exfoliated from a self-flux-grown crystal and transferred onto an oxidized Si substrate in an argon glove box (see details in the Method Section). A 4 nm Pt layer was then sputter-deposited on the $Fe_3GaTe_2$ as the high-conductivity spin current source of perpendicular and transverse spins (Fig. 1a) after it was transferred instantly into an ultrahigh vacuum sputtering system. Without the Pt layer our $Fe_3GaTe_2$ layers[12] show no current switching (see Sec. 1 in the Supplementary Information) likely due to the lack of self-generated torque and inversion symmetry breaking,[31] despite the previous reports of current switching of some $Fe_3GaTe_2$ single layers.[32-34] As shown by the optical microscopy images in Fig. 1b-d, to define the electric current and field distributions, the $Fe_3GaTe_2$ 37/Pt 4 bilayer was patterned into double-cross Hall bar device with total length of 10 μm and width of 2 μm by photolithography and ion milling, followed by the deposition of Ti 5/Pt 150 electrodes.

According to the symmetry analysis[35,36], the generation of perpendicular spins and perpendicular effective field is allowed in the normal metal/ferromagnet bilayers when the inversion symmetries are simultaneously broken in the directions that are transverse and perpendicular to the current. For our thin-film bilayer devices with thick electrodes, the electric inversion symmetry is always broken in the perpendicular direction (Fig. 1e), while the electric inversion symmetry in the transverse direction is preserved in the symmetric geometry (Fig. 1b) but not in the asymmetric geometries (Fig. 1c,d). As indicated schematically by the curved lines in Fig. 1b-d, the transverse inversion symmetry (relative gradient) for the electric field is preserved in Fig. 1b but broken oppositely for the left and right crosses in Fig. 1c,d (asymmetric geometries) due to the left-handed and right-handed curving of current path from the electrode-$Fe_3GaTe_2$/Pt-electrode. As well-established in refs. 37-40 and widely accepted in the community[41-46], it is a trivial mechanism that current flow in the ferromagnet/Pt layers in the asymmetric geometries in Fig. 1c,d additionally breaks the electric asymmetry transverse to the current and thus generates perpendicular spins. The perpendicular spins can efficiently switch the PMA magnets via anti-damping torque and functions as an effective perpendicular magnetic field ($H_{\text{eff}}^{z}$) during the magnetization switching.

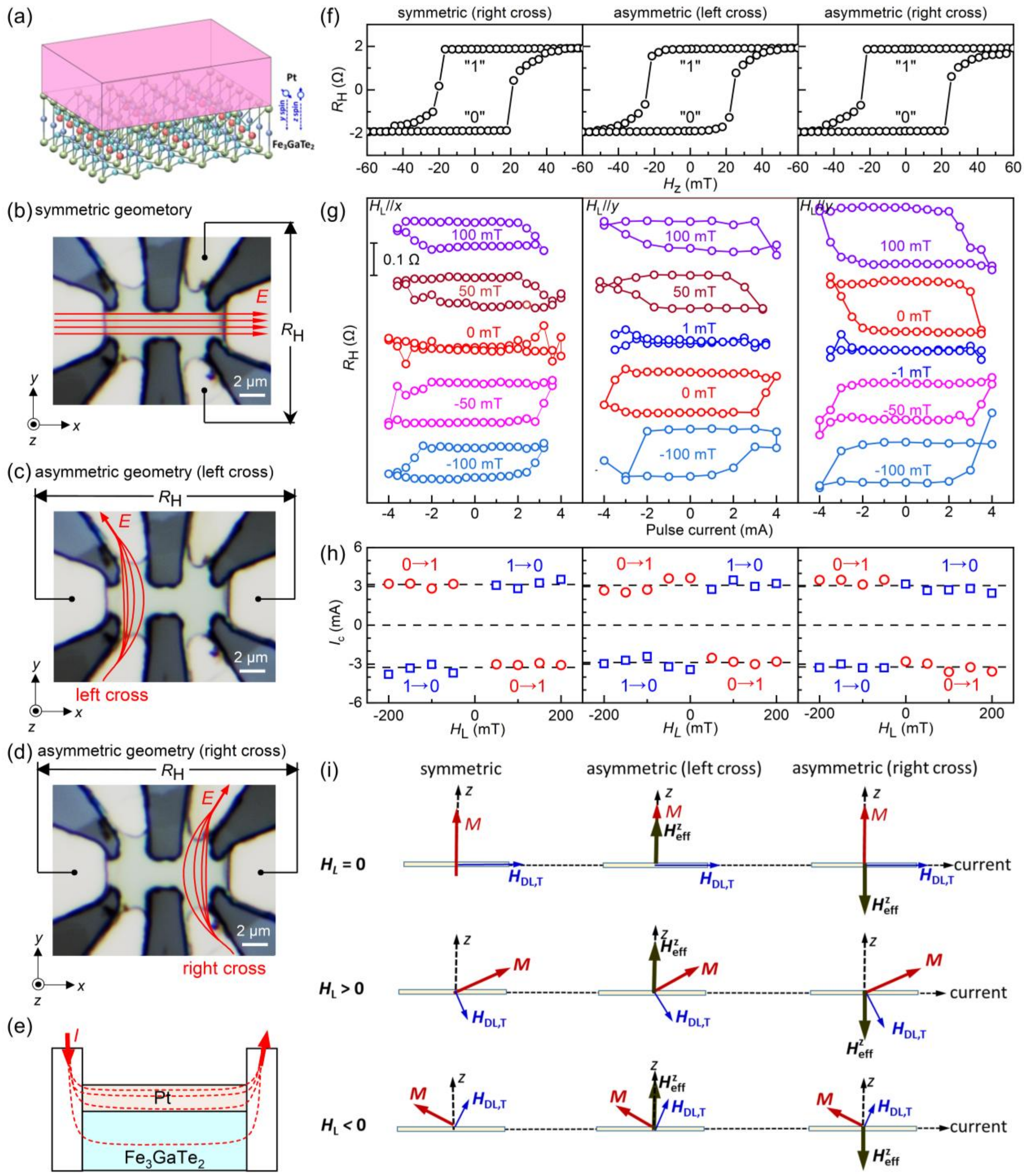

**Figure 1| Device and magnetic-field-free switching.** (a) Schematic of the $Fe_3GaTe_2$/Pt device. Optical microscopy images of the Hall bar device in (b) the symmetric geometry, (c) asymmetric geometry at the left cross, and (d) asymmetric geometry at the right cross. In (b)-(d) the red curves denote the gradient electric field distributions in the direction transverse to the current. (e) Gradient electric field distributions in the perpendicular direction. (f) Magnetic field driven Hall resistance hysteresis, (g) Current-driven Hall resistance hysteresis under different in-plane field along the current direction ($H_L$), (h) Dependence on $H_L$ of the switching current density for the right cross in the symmetric geometry, left and right crosses in the asymmetric geometries. (i) Schematic of the interplay of the perpendicular effective field of perpendicular spins ($H_{\text{eff}}^{z}$) and the dampinglike SOT field of the transverse spins ($H_{\text{DL,T}}$) under different $H_L$.

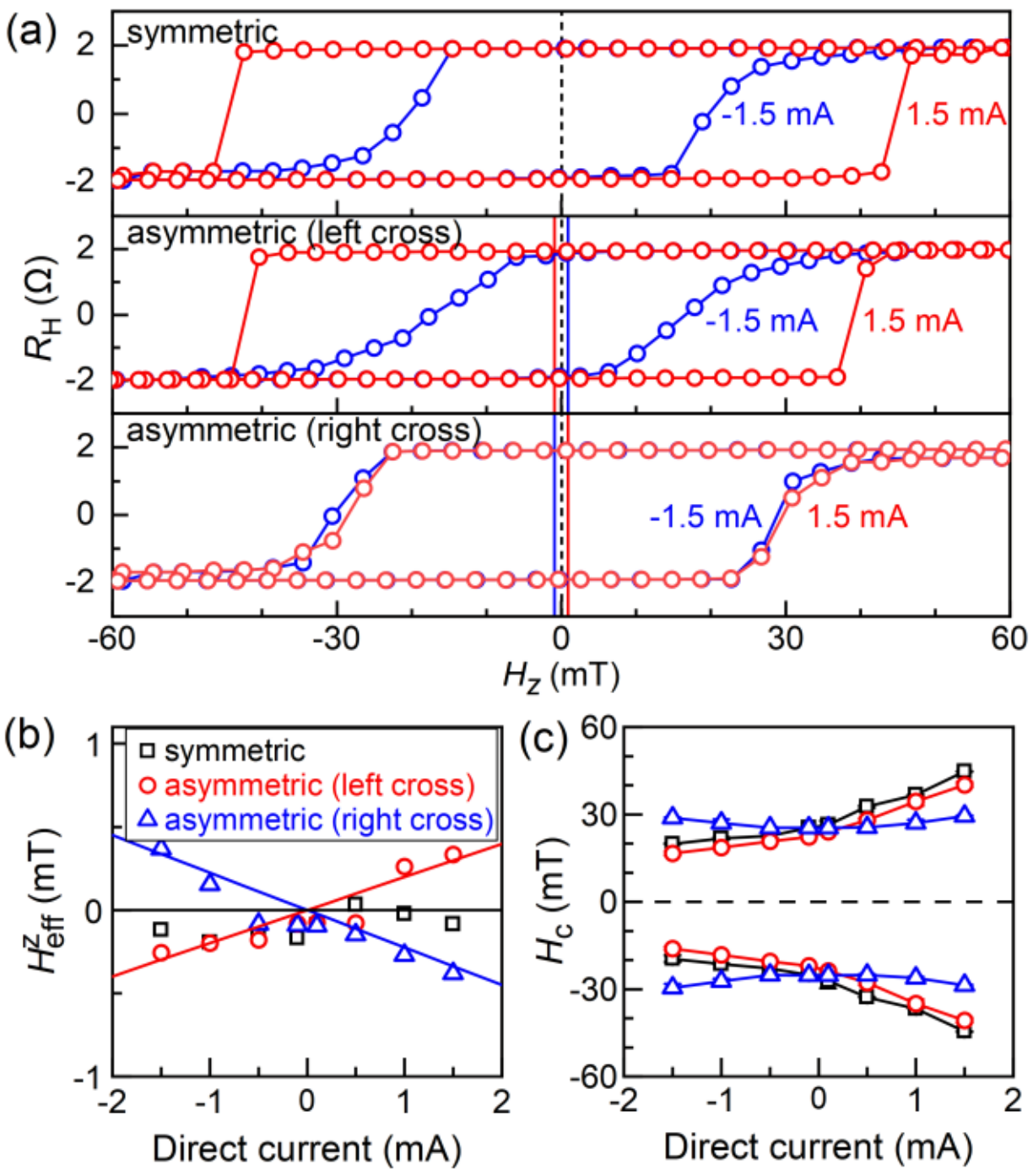

**Figure 2| Current-induced perpendicular effective field under zero in-plane magnetic field**. (a) Field-driven Hall resistance hysteresis in the symmetric and asymmetric geometries under current bias of ±1.5 mA. (b) Linear dependence on direct current of $H^z_{\text{eff}}$. (c) Variation of the coercivity with the direct current.

As shown in Fig. 1f, the device exhibits strong PMA, high coercivity of 20 mT, and clear “1” and “0” states at room temperature as indicated by the Hall resistance ($R_H$) hysteresis driven by the perpendicular magnetic field ($H_z$). From Fig. 1g, the device exhibits current switching only when an in-plane field is applied in the symmetric geometry, for which the current flows symmetrically along the 10 μm-long strip (in the $x$ direction) of the Hall device (Fig. 1b). In contrast, the device exhibits magnetic-field-free current switching in the asymmetric geometries by asymmetrically sourcing a transverse current in the transverse direction into the left or right cross (Fig. 1c,d). As expected, the switching for the left and right crosses are of opposite polarities at zero magnetic field and absent at in-plane longitudinal magnetic field ($H_L$) of +1 mT and -1 mT, respectively. Under sufficiently high $H_L$ beyond ±10 mT, the device shows the same switching polarity for all the geometries, suggesting a switching dominated by the effective SOT field of the transverse spins ($H_{DL,T}$). As shown in Fig. 1h, the switching current is about 3.0 mA (current density of $1.3\times10^7$ A/cm$^2$) for the 200 μs pulse width ($\tau$) and exhibits only a weak variation as a function of $H_L$ for both the symmetric and asymmetric geometries.

The distinct current switching characteristics in the three geometries can be phenomenologically understood by the toy model schematically depicted in Fig. 1i that considers the interplay of the perpendicular spins ($H^z_{\text{eff}}$), the transverse spins ($H_{DL,T}$), and the applied longitudinal magnetic field ($H_L$). For $H_L$ = 0, the in-plane current generates an in-plane $H_{DL,T}$ for all the geometries and a $H^z_{\text{eff}}$ of opposite signs for the left and right asymmetric crosses. In this case, the PMA device exhibits $H^z_{\text{eff}}$ -driven magnetic-field-free current switching of anticlockwise (clockwise) polarity in the left (right) cross in the asymmetric geometry but no current-induced switching in the symmetric geometry. Note that nucleation-dominated perpendicular magnetization switching[47-49] requires a perpendicular effective field component greater than the nucleation field of the device. Under $H_L > 0$ ($H_L < 0$), the magnetization is tilted towards the film plane, and $H_{DL,T}$ has a large perpendicular component that can subtract from (add to) $H^z_{\text{eff}}$ in the left cross but add to (subtract from) $H^z_{\text{eff}}$ in the right cross. In this case, all the three geometries may exhibit current switching except for the $H_L$ values at which $H^z_{\text{eff}}$ and the perpendicular component of $H_{DL,T}$ are cancelled out (i.e., $H_L$ = + 1 mT for the left cross and -1 mT for the right cross in Fig. 1g). For very high $H_L$ values, the perpendicular component of $H_{DL,T}$ becomes large and dominates the switching. The insensitivity of the switching current to $H_L$ in Fig. 1h is similar to that of many metallic SOT heterostructures[50,51] and reminiscent of previous reports of complicated, non-monotonic tuning of the nucleation field (coercivity) of metallic PMA magnets by in-plane magnetic field.[8,47,48] A more rigid theory considering the ultrafast micromagnetic dynamics would be more informative but beyond the scope of this work.

We further quantify the perpendicular effective magnetic field that drives the magnetic field-free switching in the asymmetric geometries from direct current ($I$)-induced loop shift[52] *in the absence of any applied magnetic field*. During the loop shift measurement, $R_H$ is recorded as a function swept $H_z$ while continuously applying a constant $I$. As shown in Fig. 2a, the $H_z$-driven $R_H$ hysteresis exhibits no shift in the symmetric geometry but it is shifted to the left (right) side for the left cross and to the right (left) side for the right cross when an in-plane current of +1.5 (-1.5) mA is injected in the asymmetric geometry. This observation reveals the presence of current-induced $H^z_{\text{eff}}$ in the asymmetric geometries due to perpendicular spins.[37-40] As summarized in Fig. 2b, $H^z_{\text{eff}}$ is a linear function of $I$ within the experimental uncertainty, with a slope ($dH^z_{\text{eff}}/dI$) of zero for the symmetric geometry and of 0.20±0.03 (0.23±0.03) mT/mA for the left (right) cross in the asymmetric geometries. The measurement of $H^z_{\text{eff}}$ is not influenced by the intralayer Dzyaloshinskii-Moriya interaction field[8] because the latter is independent of the current and remains negligible under zero $H_L$. Such current-induced $H^z_{\text{eff}}$ explains the unnecessity of the in-plane assisting magnetic field during the SOT switching of the $Fe_3GaTe_2$/Pt device in the asymmetric geometries. The slight difference of the slopes for the left and right crosses is attributed to a small variation of the vdw ferromagnet $Fe_3GaTe_2$ during device fabrication.

We have also observed a strong dependence on the direct current of the coercivity (after extraction of the shift), which is reaffirmed by repeated measurements. As plotted in Fig. 2c, the coercivity ($H_c$) of the device coincides reasonably at zero current but varies with the current in different manners depending on the geometries. Such interesting current dependence of the coercivity is not specific to the $Fe_3GaTe_2$ in work but also observed in other $Fe_3GaTe_2$[13] and $Co_3Sn_2S_2$ flakes[53]. Here, the $H^z_{\text{eff}}$ and $H_c$ of the vdw device are measured for relatively small direct currents of $|I| \leqslant 1.5$ mA to avoid damage by the continuous

Joule heating of the direct currents. The switching current of ⩽4 mA for the device (Fig. 1e,f) suggests a perpendicular effective field of no more than 0.9 mT, which is not greater than $H_c$ at the small currents of ⩽ 1.5 mA in Fig.2c. This most likely suggests that $H_c$ under the large pulsed current of 4 mA is substantially lowered as reported previously in Refs. 13 and 53.

**Boolean logic operations.** We demonstrate the room-temperature field-free, low-power AND and NAND logic gate operations utilizing the left and right crosses of the $Fe_3GaTe_2$/Pt device in the asymmetric geometries. As shown in Fig. 3a,b, the device exhibits an anti-clockwise switching polarity at the left cross but a clock-wise switching at the right cross, with a current of < 4 mA for pulse width of 200 µs. For the logic operation, we define the positive (negative) state of the anomalous Hall resistance as the logic output "1" ("0") and define pulse current of +2.15 mA as "1" and 0 mA as "0" for both A and B inputs. The idle state of the AND (NAND) gate is initialized to "0" ("1") by a current of -4 (+4) mA. As indicated by the experiment, the left device functions as a AND gate and returns "1" only when A = B = "1", while the right cross functions as a NAND gate and returns 1 unless A = B = "1".

To evaluate the energy efficiency and thermal stability of the $Fe_3GaTe_2$/Pt logic, the total switching current ($I_c$) is measured at different pulse width ($\tau$) in the absence of magnetic field (Fig. 3c). As summarized in Fig. 3d, the switching current increases logarithmically with shortening pulse width, fit of which to the equation[28]

$$I_c = I_{c0}(1-\Delta^{-1}\ln(\tau/\tau_0)), \quad (1)$$

yields the intrinsic critical switching current ($I_{c0}$) of 4.6 ±1.4 (4.0±0.9) mA and $\Delta$ of 47±5 (50±5) for the left (right) cross. Here, $I_{c0}$ represents the intrinsic switching current in the absence of thermal fluctuations and thus always greater than the dc switching current with thermal assistance, while $\tau_0$ is the thermal attempt time which is typically 1 ns[54,55]. Finite-element analyses of current distributions within the $Fe_3GaTe_2$ 37/Pt 4 device in Fig. 4a suggests the current in the Pt layer is approximately 37.6% of the total write current in both the symmetric and asymmetric geometries. Thus, the average intrinsic switching current density $j_{c0}$ is estimated from the $I_{c0}$ values as (2.0 ± 0.6) ×$10^7$ A/cm² and (1.8 ± 0.4)×$10^7$ A/cm² for the Pt within the left and right cross areas, respectively. We note that this represents the first report of $j_{c0}$ and $\Delta$ for a uniform vdw magnet, especially for the magnetic-field-free switching of $Fe_3GaTe_2$. The slight differences in $j_{c0}$ and $\Delta$ for the left and right crosses arise likely from a small variation of the van der Waals $Fe_3GaTe_2$ layer during device fabrication. $j_{c0}$ is slightly greater for the left cross than for the right cross, which is consistent with the fact that $dH_{\mathrm{eff}}^z/dI$ is slightly smaller for the left cross than for the right cross (Fig. 2b). Note that $H_{\mathrm{eff}}^z$ should play a critical role in the magnetic-field-free switching of the PMA device.

The $j_{c0}$ value for the 37 nm thick $Fe_3GaTe_2$ is 7 times lower than that of the most optimized PMA W/CoFeB (~1nm) devices with a similar thermal stability factor ($j_{c0}$ =1.3×$10^8$ A/cm², $\Delta$ = 48).[55] Given the strong dependence of the switching current (density) on $\Delta$ and the pulse width $\tau$ (Eq. (1)), it makes little sense to compare the intrinsic switching current value of $j_{c0}$ of our device (represents the switching current density in the absence of thermal assistance) with the *thermally-assisted and in-plane field assisted* switching current densities (typically of several times $10^6$ A/cm², see Sec. 2 in the Supplementary Information) of other vdw SOT devices[13,14,20-25] with unknown $\Delta$. Note that this also represents the first report of $j_{c0}$ for a $Fe_3GaTe_2$ device.

To give a benchmark, we plot in Fig. 4b the all-electrical write power ($P$) and current channel conductivity ($\sigma_{xx}$) of different magnetic-field-free perpendicular SOT devices with a nanodot bit sitting on an industrial-level current channel with length ($L$) of 100 nm and width ($w$) of 50 nm (width). $P$ at the pulse width $\tau$ is estimated as

$$P = (I_c/w_0)^2 Lw\tau/\sigma_{xx}d, \quad (2)$$

where $w_0$ and $d$ are the original width and thickness of the SOT devices. The $Fe_3GaTe_2$ 37/Pt 4 device exhibits the highest channel conductivity of 3.3×$10^6$ Ω$^{-1}$ m$^{-1}$ and the lowest write power of 460 pJ at $\tau$ = 400 µs and 1.5 fJ at $\tau$ = $\tau_0$=1 ns. This power is about 40 times less than that of the most optimized PMA W/CoFeB (1nm) devices with a similar thermal stability factor and device dimensions.[55] This power reduction can be understood by at least three reasons: (*i*) Pt is more energy-efficient than W[56] since Pt has the much higher spin Hall conductivity (1.6×$10^6$ Ω$^{-1}$ m$^{-1}$ for Pt[57] and 1.6×$10^5$ Ω$^{-1}$ m$^{-1}$ for W[56]) and conductivity than W ( 3.3×$10^6$ Ω$^{-1}$ m$^{-1}$ for Pt and 3×$10^5$ Ω$^{-1}$ m$^{-1}$ for W); (*ii*) the $Fe_3GaTe_2$ has much lower saturation magnetization than the FeCoB (200 emu/cm$^3$ for $Fe_3GaTe_2$ and 1300 emu/cm$^3$ for $Fe_{60}Co_{20}B_{20}$)[15,57]; (*iii*) $z$ spins (via antidamping torque switching) are more energy-efficient than the combination of $y$ spins and stray field of a hard magnet[55] (anti-field switching) for enabling field-free switching of perpendicular magnetization. Since the power scales parabolically with the $Fe_3GaTe_2$ thickness and the inverse damping-like SOT efficiency, further reduction of the power is expected for $Fe_3GaTe_2$ logic devices with a few-layer $Fe_3GaTe_2$ bit and a spin Hall metal with greater SOT efficiencies of $y$ and $z$ spins (such as Pt-based alloy)[50]. For the point of view of reducing the write impedance and enhancing the device endurance, the spin Hall channel of the high-conductivity Pt is also highly preferred than low-conductivity W (3×$10^5$ Ω$^{-1}$ m$^{-1}$), PtMn (7.3×$10^5$ Ω$^{-1}$ m$^{-1}$)[14], and low-symmetry crystals of $TaIrTe_4$ (2.4×$10^5$ Ω$^{-1}$ m$^{-1}$) and $WTe_2$ (2.4×$10^5$ Ω$^{-1}$ m$^{-1}$).[16]

(a) **AND** (left cross)

| Input A | Input B | output |
| --- | --- | --- |
| "1" (2.15 mA) | "1" (2.15 mA) | "1" |
| "1" (2.15 mA) | "0" (0 mA) | "0" |
| "0" (0 mA) | "1" (2.15 mA) | "0" |
| "0" (0 mA) | "0" (0 mA) | "0" |

(b) **NAND** (right cross)

| Input A | Input B | output |
| --- | --- | --- |
| "1" (2.15 mA) | "1" (2.15 mA) | "0" |
| "1" (2.15 mA) | "0" (0 mA) | "1" |
| "0" (0 mA) | "1" (2.15 mA) | "1" |
| "0" (0 mA) | "0" (0 mA) | "1" |

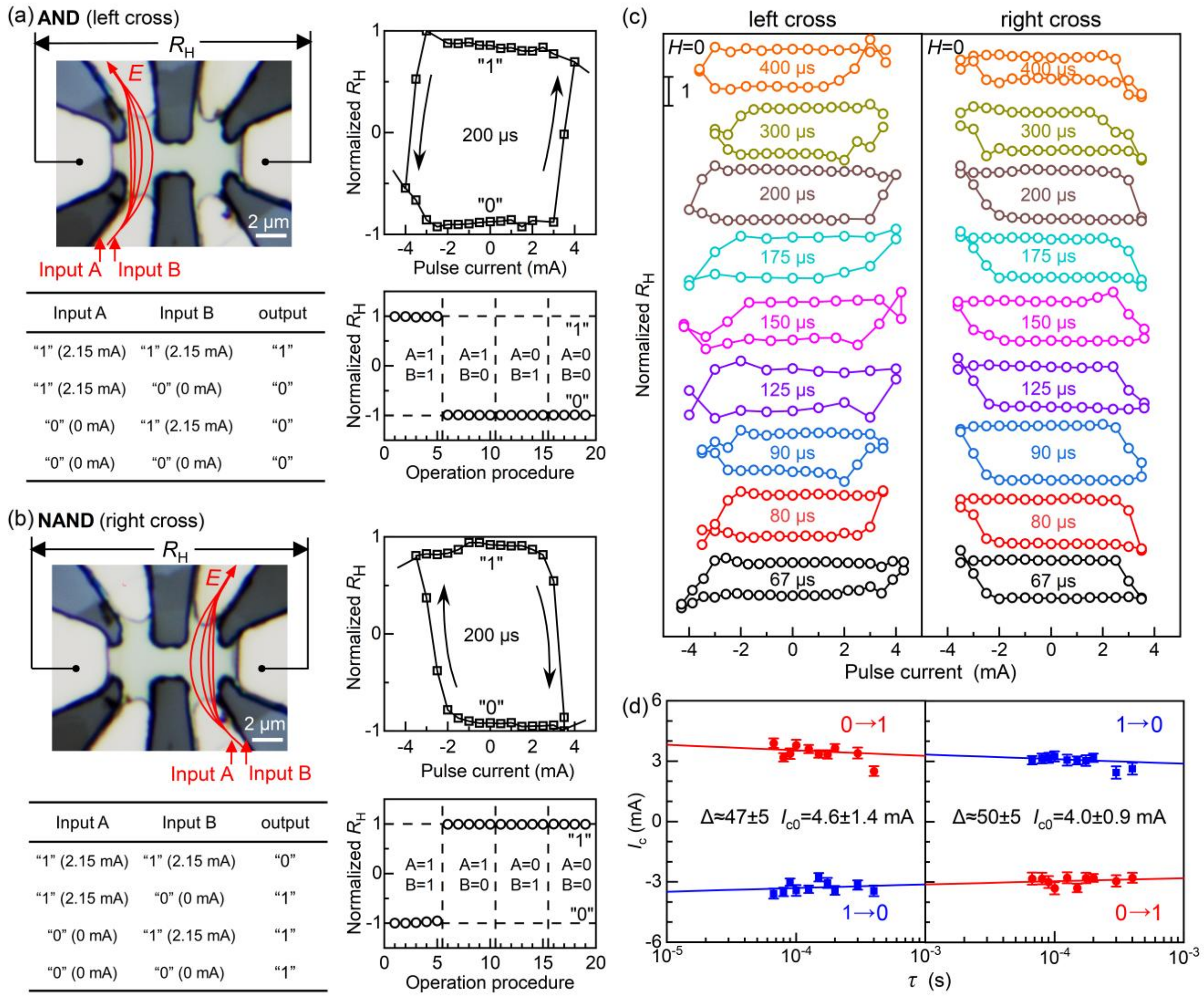

**Figure 3| Field-free low-power Boolean logic operations.** Measurement geometry, current switching ($\tau$ = 200 μs), and logic operations for (a) the AND gate (with initial state of "0") and (b) the NAND gate (with initial state of "1"). The red curves schematically denote the electric field distributions transverse to the current. Dependences on the current pulse width of (c) the anomalous Hall resistance hysteresis and (d) the total switching current of the AND and NAND gates.

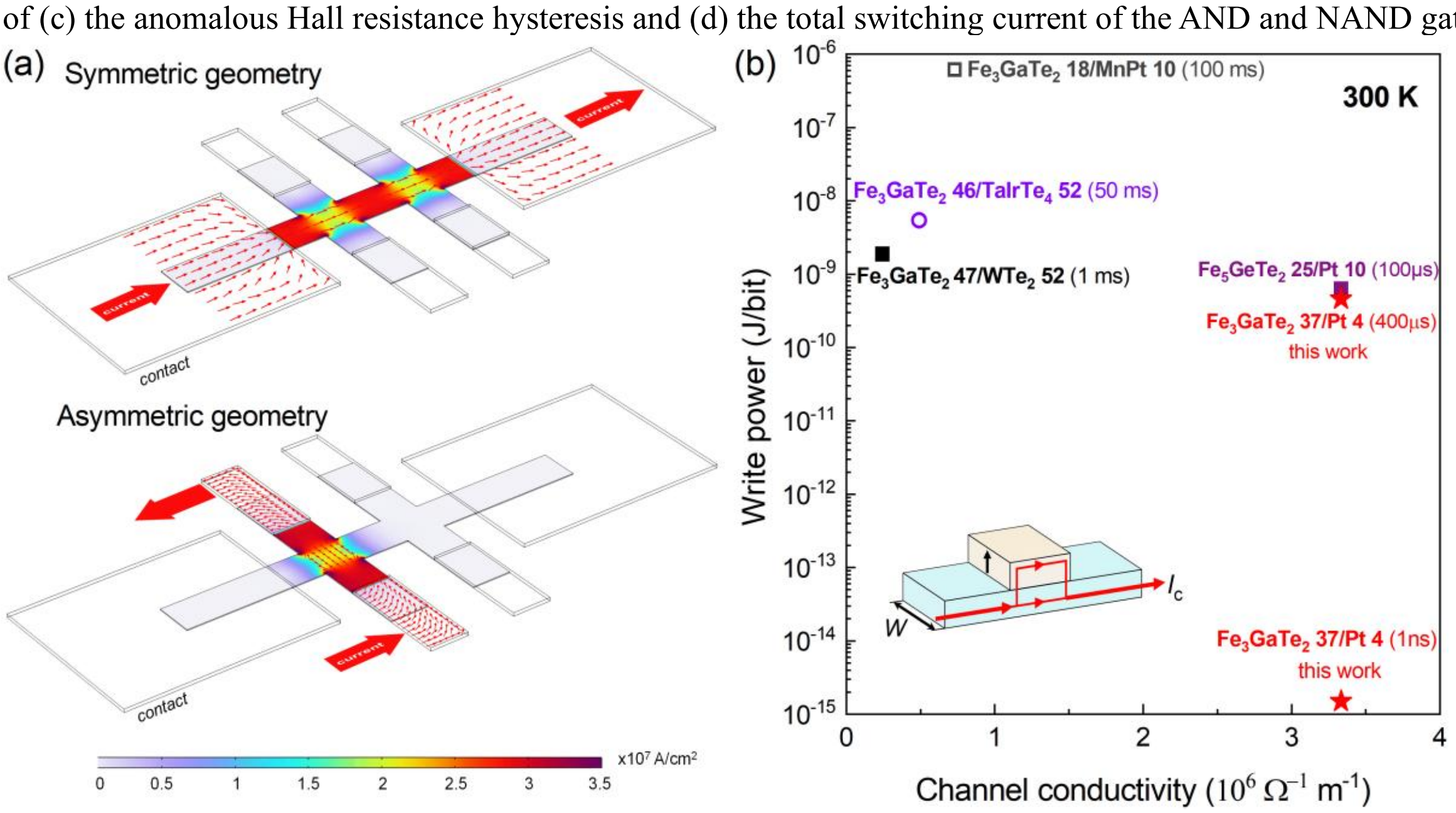

**Figure 4| Current distribution and write power.** (a) Finite-element analysis of the current distributions within the $Fe_3GaTe_2$ 37/Pt 4 Hall device in the symmetric and asymmetric geometries. (b) Comparison of the write power and the channel conductivity of the magnetic-field-free vdw SOT devices with industrial-level channel dimension of 100 nm × 50 nm. The numbers following the materials are the thicknesses in nanometers.

Note that the write powers $P$ at different pulse widths we measure here should represent the exact performance of the device in practical applications and that thermal assistance described by Eq. (1) is inevitable and beneficial. The all-electrical switching of the device is dominated by the SOT since the temperature is well below the Curie temperature of 349 K despite the current-induced Joule heating (Sec. 3 in the Supplementary Information). In addition to the AND and NAND gates, other energy-efficient logic gate functions may be developed in the future if the symmetry of the bipolar switching currents can be broken, e.g., by optimizing the Dzyaloshinskii–Moriya interaction effects[8] of the $Fe_3GaTe_2$/heavy metal bilayers. The simple device architecture also allows for high-sensitivity readout using the tunnel magnetoresistance (TMR) of a magnetic tunnel junction.

**Conclusion**: We have demonstrated a room-temperature, field-free, low-power, scalable SOT Boolean logic device architecture utilizing uniform vdw ferromagnet $Fe_3GaTe_2$ with strong perpendicular magnetic anisotropy, the strong spin Hall effect of Pt, and the perpendicular effective field from the spin Hall metal with electric asymmetries. Current pulse width-dependent measurement for the first time reveals a very low intrinsic magnetic-field-free critical current of $1.8\times10^7$ A/cm² (in the absence of thermal fluctuations and external magnetic field) and a high thermal stability factor of 50 for the $Fe_3GaTe_2$ 37/Pt 4. The switching current density of the 37 nm thick $Fe_3GaTe_2$ is already 7 times lower than that of the most optimized PMA W/CoFeB (1nm) devices with a similar thermal stability factor.[55] These results suggest the potential of an ultralow operation power of 1.5 fJ/bit at 1 ns pulse width for the $Fe_3GaTe_2$/Pt logic device with industry-level lateral dimensions, which are 40 times less than the optimized PMA W/CoFeB (1nm) devices and 3-5 orders of magnitude smaller than van der Waals semiconductor-based charge transistor logic devices (typically 1-100 pJ/bit)[58,59]. Further reduction of the power is expected for $Fe_3GaTe_2$ logic devices with a few-layer $Fe_3GaTe_2$ bit and a spin Hall channel with greater SOT efficiencies of perpendicular and transverse spins and relatively high conductivity (such as Pt-based alloy)[50]. These results highlight the great expectation of van der Waals magnet $Fe_3GaTe_2$ in dense, ultralow-power, scalable, room-temperature all-electrical in-memory computing technologies. This work will inspire in-memory computing spin logic and promote the room-temperature vdw magnets into advanced computing chip applications.

## Methods

**$Fe_3GaTe_2$ single crystal growth.** High-quality $Fe_3GaTe_2$ single crystals were grown by the self-flux method. Fe powder (99.98%), Ga chunks (99.99%), and Te chunks (99.999%) were mixed in a molar ratio of 1:1:2 and heated to 1000 °C for 48 hours in a vacuum quartz tube.

**Device fabrication.** A $Fe_3GaTe_2$ layer is first mechanically exfoliated from the single crystal and transferred onto an oxidized Si substrate in an argon glove box with the $H_2O$ and $O_2$ pressures below 0.01 ppm. A 4 nm Pt layer was then sputter-deposited on the $Fe_3GaTe_2$ as the spin current source after it was transferred instantly into an ultrahigh vacuum sputtering system with base pressure of ~$10^{-10}$ Torr. The layer thickness of $Fe_3GaTe_2$ is measured to be 37 nm by a Bruker atomic force microscopy. The $Fe_3GaTe_2$/Pt bilayer was patterned into double-cross Hall bar device by ultraviolet photolithography and ion milling. Finally, electrical electrodes of Ti(5nm)/Pt(150nm) were fabricated using sputter-deposition and lift off.

**Electrical measurements.** The switching and logic performance measurements were measured by a Keithley 6221 current source and Keithley 2182A nanovoltmeter.

**Finite-element analysis**. The current distribution within the $Fe_3GaTe_2$ (37 nm)/Pt (4 nm) device is simulated using finite-element analysis following the current continuity equation $\nabla\cdot\boldsymbol{j}_c+\frac{\partial\rho}{\partial t}=0$ and the boundary condition of $\boldsymbol{n}\cdot\boldsymbol{j}_c=0$, where $\boldsymbol{j}_c$ is the charge current density, $\rho$ is the charge density, $\boldsymbol{n}$ is the normal direction of the device boundary. The analysis has taken into account the large Ti/Pt electrodes and the dimensions of the real device (see Fig. 1c). The conductivity is $3.3\times10^6\ \Omega^{-1}\ m^{-1}$ for the Pt layer, $3.4\times10^5\ \Omega^{-1}\ m^{-1}$ for the $Fe_3GaTe_2$ layer, and $4.2\times10^6\ \Omega^{-1}\ m^{-1}$ for the electrode. The current within the Pt cross area is found to be about 37.6% of the injected current of the whole device.

**Acknowledgements**
This work is supported partly by the National Key Research and Development Program of China (2022YFA1204000), the Beijing Natural Science Foundation (Z230006), and the National Natural Science Foundation of China (12274405 and 12204297).

**Author contributions:** L. Z. conceived the project, Z. L. fabricated and measured the devices, Q. L. performed the finite-element analysis, W.Z. grew the $Fe_3GaTe_2$ single crystal, L. Z. and Z. L. wrote the manuscript.

**Competing interests:** The authors declare no competing interests.

**Data and Materials Availability:** All data are present in the paper.